\documentclass[aps,pra,reprint,amsmath,amssymb,superscriptaddress]{revtex4-1}

\usepackage{color}
\usepackage{graphicx}
\usepackage{dcolumn}
\usepackage{amsthm}
\usepackage{bm}
\usepackage{siunitx}
\usepackage{nicefrac}
\usepackage{todonotes}
\usepackage{mathtools}
\usepackage{upgreek}
\usepackage{url}
\usepackage{physics}

\newcommand{\figref}[1]{Fig.~\ref{fig:#1}}

\newcommand{\eqnref}[1]{Eq.~(\ref{eq:#1})}

\newcommand{\AuGe}{$\mathrm{Au}_x\mathrm{Ge}_{1-x}$\,}
\newcommand{\AuGex}[2]{$\mathrm{Au}_{#1}\mathrm{Ge}_{#2}$\,}

\newcommand{\punc}[1]{\,#1}
\newcommand{\kB}{k_{\mathrm{B}}}

\begin{document}

\title{Au-Ge alloys for wide-range low-temperature on-chip thermometry}
\author{J. R. A. Dann}
\email{jrad3@cam.ac.uk}
\affiliation{Cavendish Laboratory, University of Cambridge, J.J. Thomson Avenue, Cambridge CB3 0HE, UK}
\author{P. C. Verpoort}
\affiliation{Cavendish Laboratory, University of Cambridge, J.J. Thomson Avenue, Cambridge CB3 0HE, UK}
\author{J. Ferreira de Oliveira}
\affiliation{Cavendish Laboratory, University of Cambridge, J.J. Thomson Avenue, Cambridge CB3 0HE, UK}
\affiliation{Centro Brasileiro de Pesquisas Físicas, Rua Dr Xavier Sigaud 150, Rio de Janeiro, 22290-180, Brazil}
\author{S. E. Rowley}
\affiliation{Cavendish Laboratory, University of Cambridge, J.J. Thomson Avenue, Cambridge CB3 0HE, UK}
\affiliation{Centro Brasileiro de Pesquisas Físicas, Rua Dr Xavier Sigaud 150, Rio de Janeiro, 22290-180, Brazil}
\author{A. Datta}
\affiliation{Department of Materials Science, University of Cambridge, 27 Charles Babbage Road, Cambridge CB3 0FS, UK}
\author{S. Kar-Narayan}
\affiliation{Department of Materials Science, University of Cambridge, 27 Charles Babbage Road, Cambridge CB3 0FS, UK}
\author{C. J. B. Ford}
\affiliation{Cavendish Laboratory, University of Cambridge, J.J. Thomson Avenue, Cambridge CB3 0HE, UK}
\author{G. J. Conduit}
\affiliation{Cavendish Laboratory, University of Cambridge, J.J. Thomson Avenue, Cambridge CB3 0HE, UK}
\author{V. Narayan}
\email{vn237@cam.ac.uk}
\affiliation{Cavendish Laboratory, University of Cambridge, J.J. Thomson Avenue, Cambridge CB3 0HE, UK}
\date{\today}

\begin{abstract}

We present results of a Au-Ge alloy that is useful as a resistance-based thermometer from room temperature down to at least \SI{0.2}{\kelvin}. Over a wide range, the electrical resistivity of the alloy shows a logarithmic temperature dependence, which simultaneously retains the sensitivity required for practical thermometry while also maintaining a relatively modest and easily-measurable value of resistivity. We characterize the sensitivity of the alloy as a possible thermometer and show that it compares favorably to commercially-available temperature sensors. We experimentally identify that the characteristic logarithmic temperature dependence of the alloy stems from Kondo-like behavior induced by the specific heat treatment it undergoes.

\end{abstract}

\maketitle

\section{Introduction}

Measuring the temperature ($T$) dependence of material properties is crucial in building up our understanding of condensed matter systems. It is also important for the control and measurement of electronic devices, particularly quantum devices, that operate at low $T$. Cooling materials to very low $T$ has often revealed rich and unexpected physics with relevance for current and future technologies. Understanding and quantifying low-temperature phenomena requires effective thermometry across the entire $T$ range of interest.

A limitation of present-day thermometry schemes is the absence of a single sensor spanning a broad $T$ range without loss of sensitivity. It is therefore customary to use multiple thermometers, each of which is designed for use in different parts of the $T$ range of interest. For example, in resistance ($R$)-based thermometry, in which $T$ is inferred from the electrical resistance of a `thermistor', metallic or semiconducting sensors are used depending on the $T$ range in question: high $T$ is most conveniently measured using metallic thermistors which, however, lose sensitivity below $\approx \SI{5}{\kelvin}$, where $R$ is dominated by $T$-independent impurity scattering. At low $T$, semiconducting thermistors (e.g., Ge, RuO$_2$, carbon glasses) are employed, in which the $R$ increases as $T$ is lowered with activated behavior, $R \sim \exp(\Delta/\kB T)$. Yet despite the large sensitivity that such thermistors offer, the rapidly increasing $R$ can raise substantial issues with measurements that avoid self-heating of the thermometer.

A variety of other techniques are also available for thermometry at low $T$, such as nuclear orientation thermometry~\cite{marshak1983nuclear}, magnetic thermometry~\cite{schuster1994thermometry}, shot-noise thermometry~\cite{spietz2003shot-noise}, Johnson-noise thermometry~\cite{white1996johnson-noise}, Coulomb-Blockade thermometry~\cite{pekola1994coulomb-blockade, kauppinen1998coulomb}, capacitance thermometry~\cite{lawless1971capacitance}, lithographically-defined on-chip thermocouples~\cite{chickering2009thermocouple, narayan2012thermocouple, billiald2015thermocouple, narayan2014thermopower}, non-invasive charge-sensing thermometers~\cite{mavalankar2013charge-sensing}, and normal-superconductor-based tunnel junctions~\cite{feshchenko2015tunnel, saira2016tunnel}.

\begin{figure}
    \includegraphics[width=\columnwidth]{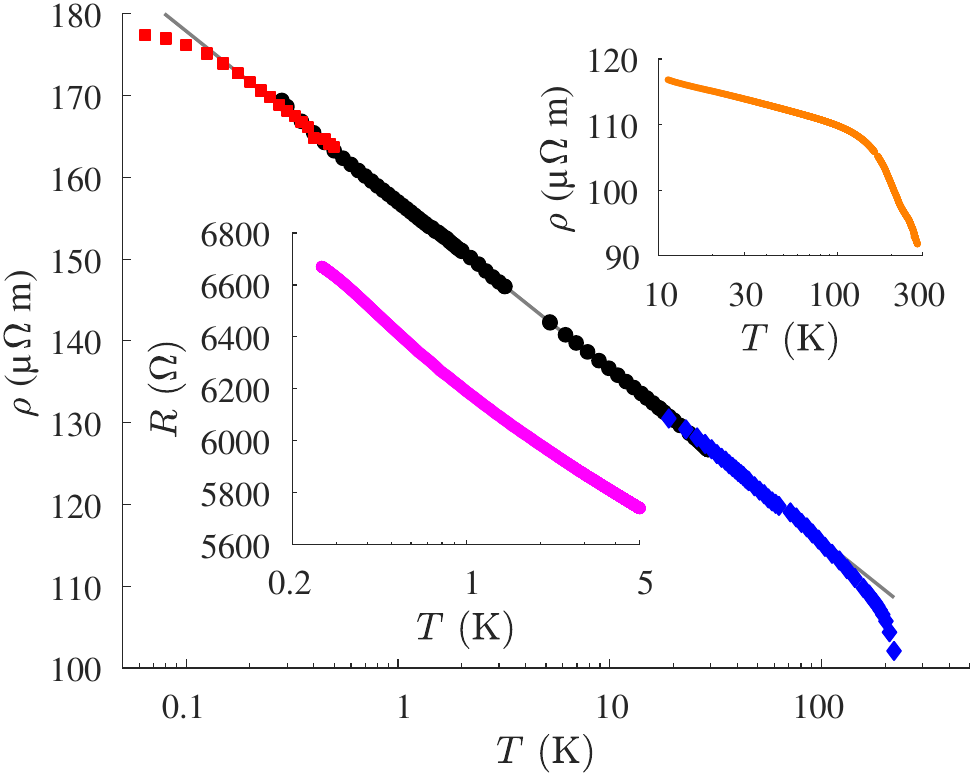}
    \caption{Resistivity ($\rho$) vs temperature ($T$) for a \AuGe film ($x = 0.07$). The \AuGe film shows a monotonic decrease in $\rho$ from \SI{0.2}{\kelvin} all the way up to room $T$, thus behaving as a thermistor over a very wide $T$ range. The behavior up to $\approx\SI{200}{\kelvin}$ is described well by a logarithmic equation as explained in the main text. Different markers correspond to measurements taken in different cryostats or while cooling, normalized by multiplicative factors of 1-2\%. The solid line shows a fit to the data shown in black circles. The top inset (orange curve) shows measurements on another \AuGe sample ($x \approx 0.1$), taken while cooling from room temperature. The bottom inset (magenta curve) shows $R \left(T\right)$ measurements for a \AuGe ($x = 0.07$) film grown on a sapphire substrate, taken in a third cryostat while $T$ was ramped continuously. The \AuGe alloy can be used as a single sensor operable between \SI{0.2}{\kelvin} and \SI{200}{\kelvin}.}
    \label{fig:AG3-1a_RvsT}
\end{figure}

Here we present a novel thermistor which offers several of the individual advantages presented by these techniques, namely: 1) it is sensitive over a broad $T$-range; 2) its resistance remains easily measurable and small enough to allow efficient thermalization and avoid self-heating; 3) it can be lithographically patterned to reliably measure spatial $T$ gradients and/or be fabricated on chips as part of microelectronic devices.

The thermistor we report is made of a Au-Ge alloy (\AuGe) and might be suitable as a $R$-based thermometer that can operate over a wide $T$ range from \SI{0.2}{\kelvin} up to \SI{200}{\kelvin}, and possibly even up to room $T$. In particular, we demonstrate that for a range of compositions $R$ shows a logarithmic dependence on $T$ across a range of three orders of magnitude. The material satisfies the conditions: 1) the sensitivity $S$ remains usably high between room $T$ and \SI{0.2}{\kelvin} with $S$ increasing with lowering $T$, 2) the absolute value of $R\left(T\right)$ remains sufficiently small that self-heating should be negligible, and thermalization should not be an issue, 3) this material is suitable for lithographic \si{\micro\meter}-scale fabrication, and thus for use on chips as part of nano-scale devices, as well as on larger scales for use in other environments, such as cryogenic refrigerators, and, in addition, 4) the linearity of $R$ as a function of $\log \left(T\right)$ allows for a relatively simple calibration.

\section{Methods}

The \AuGe alloy was fabricated into Hall bar devices using a photolithography and lift-off technique. The required thicknesses of Ge and Au were thermally evaporated in separate layers and the Hall bars were then heated in a reducing environment of nitrogen and hydrogen at $\approx \SI{450}{\celsius}$ for two minutes before being rapidly cooled to room $T$ within a few minutes, as shown in Fig.~S1. Finally, Ohmic Ti/Au contacts were deposited in a second photolithography step. The composition of the material was measured in two separate ways: 1) the thickness of each layer was measured using a crystal monitor during the evaporation; and 2) energy dispersive X-ray spectroscopy measurements (see Fig.~S2) were carried out after the heat treatment. The compositions from the two methods were consistent within 0.2\%. The Hall bars were approximately \SI{1}{\milli\metre} long, \SI{80}{\micro\metre} wide, and \SI{200}{\nano\metre} thick. Several different substrates have been successfully employed, namely silicon, sapphire, and GaAs. Unless otherwise stated, all measurements presented here were carried out on samples deposited on silicon substrates. The results are not observed to be affected by the choice of three substrates, and showed the same main characteristics over multiple separate processing batches. Resistance measurements were carried out using standard four-terminal low-frequency lock-in techniques. Unless otherwise stated, a frequency of \SI{77}{\hertz} was used and the excitation current was \SIrange{1}{100}{\nano\ampere}. The measurements reported here were performed in three different cryostats with base temperatures of \SI{0.05}{\kelvin}, \SI{0.2}{\kelvin}, and \SI{0.3}{\kelvin}. While the qualitative nature of the results was unaffected between cryostats, in order to make quantitative comparisons of the results (\figref{AG3-1a_RvsT}), we have introduced multiplicative factors of 1-2\% in \figref{AG3-1a_RvsT} to ensure that data from different cryostats falls on the same curve.

\section{Results and Discussion}

Figure~\ref{fig:AG3-1a_RvsT} shows the $T$ characteristics of a \AuGex{0.07}{0.93} film. Over most of the $T$ range, the electrical resistivity $\rho \equiv R \times Wt/L$ (here $W$ and $L$ are the width and length of the Hall bar respectively, and $t$ is the film thickness) displays a logarithmic $T$ dependence increasing by $\approx \SI{30}{\micro\ohm\meter}$ per decade in $T$. The data for the main panel of this figure has been taken in two separate cryostats to span the full $T$ range. Notably, $\rho$ changes monotonically from room $T$ down to $T^{*} \approx$~\SI{0.2}{\kelvin}, below which the dependence begins to flatten out. Above $T^{*}$, $\mathrm{d} \rho / \mathrm{d} \log T$ is constant (and so $\rho \left( \log T \right)$ is linear) until $\approx$~\SI{200}{\kelvin}, where there is a distinct change in slope. Between these limits the data is fitted to

\begin{equation}
\rho \left(T\right) = - \rho_0 \log\left(\frac{T}{T_0}\right), \label{eq:rho_log}
\end{equation}
where $\rho_0$ and $T_0$ are fitting parameters depending on the exact composition of the alloy.

Not only is the logarithmic form of $\rho \left(T\right)$ applicable over a wide $T$ range, but $\rho$ also shows little deviation from this functional form. This accuracy in the form of $\rho \left(T\right)$ should make this material useful for $R$-based thermometry, as any deviations adversely impact the accuracy of the measured $T$. Furthermore, the simplicity and accuracy of \eqnref{rho_log} makes calibration straightforward. Calibrating a \AuGex{0.07}{0.93} thermometer would only require measurements at two well-defined values of $T$, for example when immersed in liquid nitrogen and in liquid helium. More calibration points could improve the calibration and/or extend the $T$ range over which it is calibrated.

The relatively straightforward calibration offsets an unexpected cooldown dependence of R. As shown in Fig. S3, we observe a non-systematic variation of R between cooldowns. While this effect is small, it is certainly within experimental resolution. The origin of this effect is unexplained as of now, and may have to do with the differential thermal expansion of the film with respect to the substrate. However, we can rule out the role of atomic diffusion as the resistance of the samples is stable when stored at room temperature for several weeks or months. Likewise, the resistance does not vary when maintained at sub-Kelvin temperatures, implicating the temperature change as responsible for the observed cooldown dependence. We have noticed that encapsulating the film in a few hundred \si{\nano\metre} of Ge substantially improves the stability. However, even in the absence of capping, a simple two-point calibration should mitigate the error introduced by this variation.

\begin{figure}
   \includegraphics[width=\columnwidth]{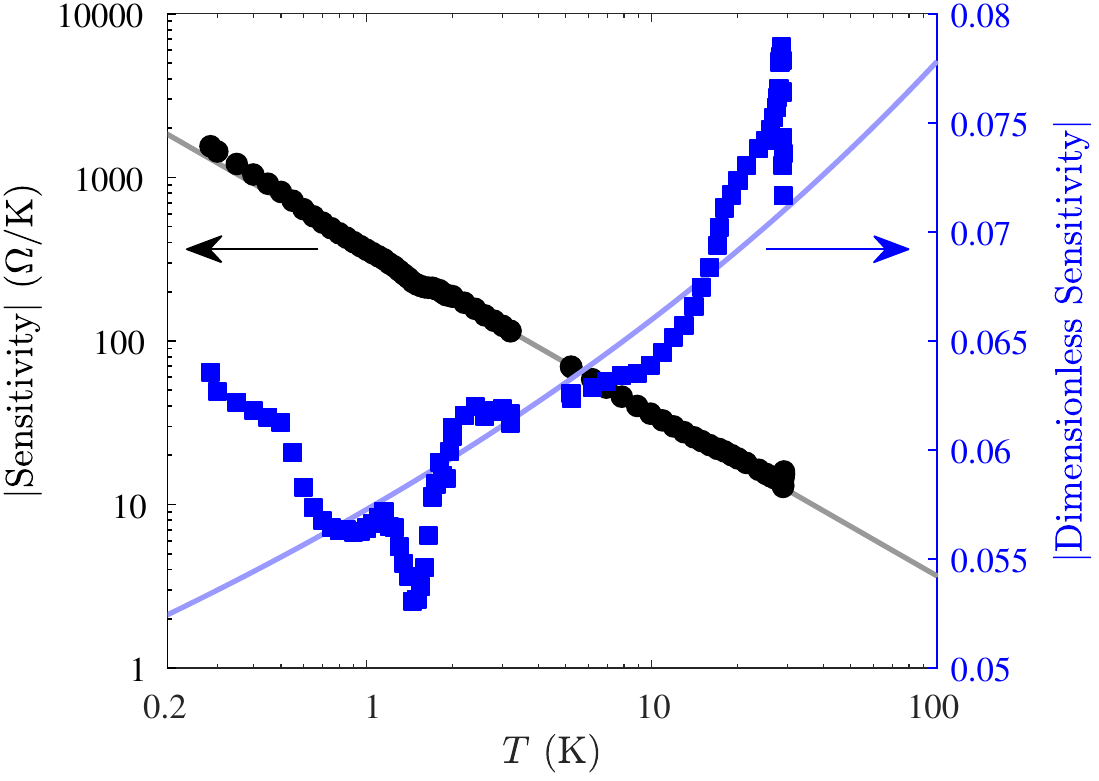}
    \caption{Sensitivity $S \equiv \dv*{R}{T}$ (left axis, black circles) and dimensionless sensitivity $\frac{T}{R} \dv{R}{T}$ (right axis, blue squares) for the \AuGex{0.07}{0.93} sample. The $1/T$ dependence (solid line) of $S$ is consistent with the logarithmic behavior of $R(T)$. The dimensionless sensitivity varies by only $\approx \SI{60}{\percent}$ over nearly three orders of magnitude in $T$. Here the points have been extracted by numerically differentiating the data, whereas the solid lines have been calculated from the fit shown in \figref{AG3-1a_RvsT}.}
    \label{fig:AG3-1a_Sensitivity}
\end{figure}

Figure~\ref{fig:AG3-1a_Sensitivity} shows that down to $T \approx \SI{0.2}{\kelvin}$ the sensitivity $S \equiv \dv*{R}{T}$ of the \AuGex{0.07}{0.93} alloy as a thermometer increases as $T \rightarrow \SI{0}{\kelvin}$. Although this dependence is found even in semiconductor materials, in those the advantage gained due to the increasing $S$ is offset by the correspondingly large values of $R$. This is the primary difference between semiconductors and the \AuGe alloy, in which the absolute value of $R$ remains moderate. Therefore, we examine not only $S$ but also the `dimensionless sensitivity' $\frac{T}{R} \frac{dR}{dT}$ which, importantly, is found to vary by less than a factor of 2 over three decades in $T$. This is to be contrasted with commercially-available thermometers, in which the dimensionless sensitivity typically varies by more than an order of magnitude over comparable $T$ scales, restricting the $T$ range over which they can be used~\cite{lakeshore-cernox}. The weak dependence of the dimensionless sensitivity on $T$ suggests not only that a single sensor element can be used over a wide $T$ range, but also that the same electronic instrument required to measure $R$ can be used over the whole $T$ range.

\begin{figure}
    \includegraphics[width=\columnwidth]{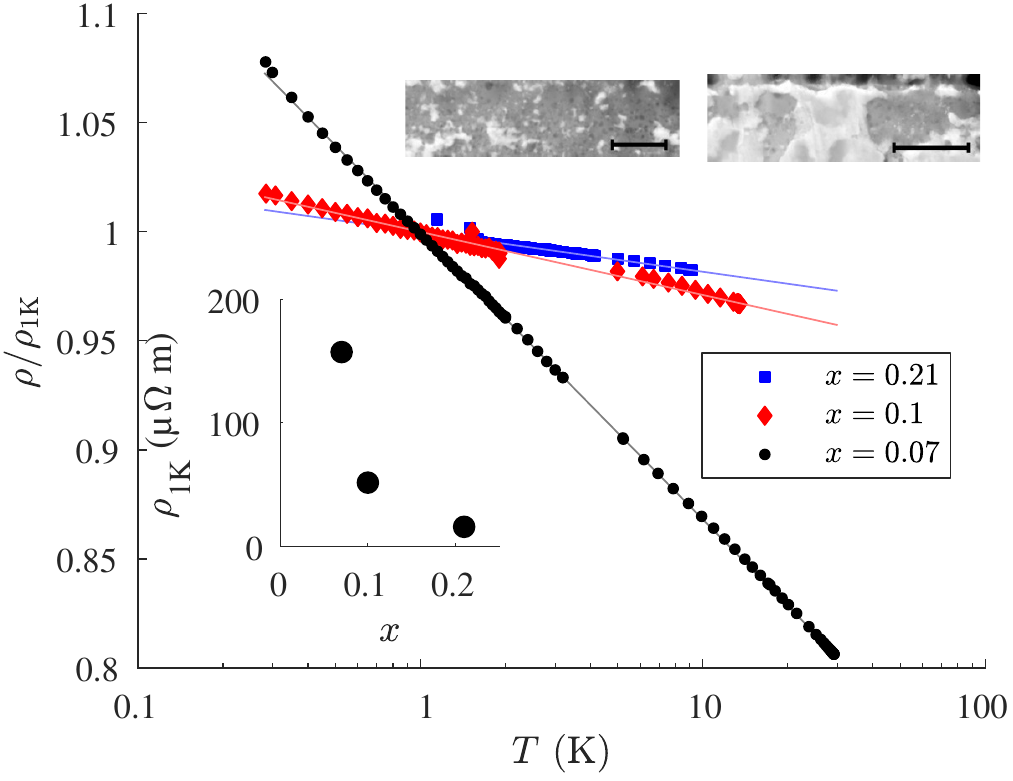}
    \caption{$\rho(T)$ for three \AuGe compositions. The main panel shows $\rho(T)/\rho(\SI{1}{\kelvin})$ for different compositions and the solid lines show the respective logarithmic fits. The bottom inset shows $\rho(x)$ at \SI{1}{\kelvin}. The top insets show cross-sectional TEM images of \AuGex{0.07}{0.93} (left) and \AuGex{0.21}{0.79} (right) in which the brighter regions are more Au-rich. The scale bar in each represents \SI{200}{\nano\metre}. The resistivity as measured between different pairs of contacts is consistent suggesting that the disorder that exists on short lengthscales is averaged out in the macroscopically large devices studied here.}
    \label{fig:RvsTAllSamples}
\end{figure}

It is instructive to consider how the fractional error in a $T$ reading varies with $T$. Normally, the dimensionless sensitivity gives an indication of this, assuming that the fractional error in the resistance measurement remains constant throughout the $T$ range of operation. However, since $\rho$ of the Au-Ge film changes only relatively slowly with $T$, changing by less than one order of magnitude over the entire $T$ range studied, it is reasonable to consider the case of resistance measurements having a fixed uncertainty $\sigma_{R}$ over the entire temperature range. Using \eqnref{rho_log}, this leads to an uncertainty in the measured $T$ of $\sigma_T \sim T \sigma_R$. Over the entire range of applicability of \eqnref{rho_log} the fractional uncertainty in the $T$ measurement is therefore constant. Therefore, the Au-Ge alloys described here have the potential to be used to make thermometers that are equally useful at low and relatively high $T$.

We now explore how changing the composition affects the behavior of the material. A similar logarithmic $\rho \left(T\right)$ dependence has been observed in samples with a variety of compositions of \AuGe ($0.07 \leq x \leq 0.21$). Figure~\ref{fig:RvsTAllSamples} shows that $S$ is the largest for $x = 0.07$, for which the uncertainties in the fitting parameters are also smallest. While this suggests that both $S$ and the accuracy may be increased by further reducing $x$, we also note that this would increase the absolute value of $\rho$.

\begin{figure}
    \includegraphics[width=\columnwidth]{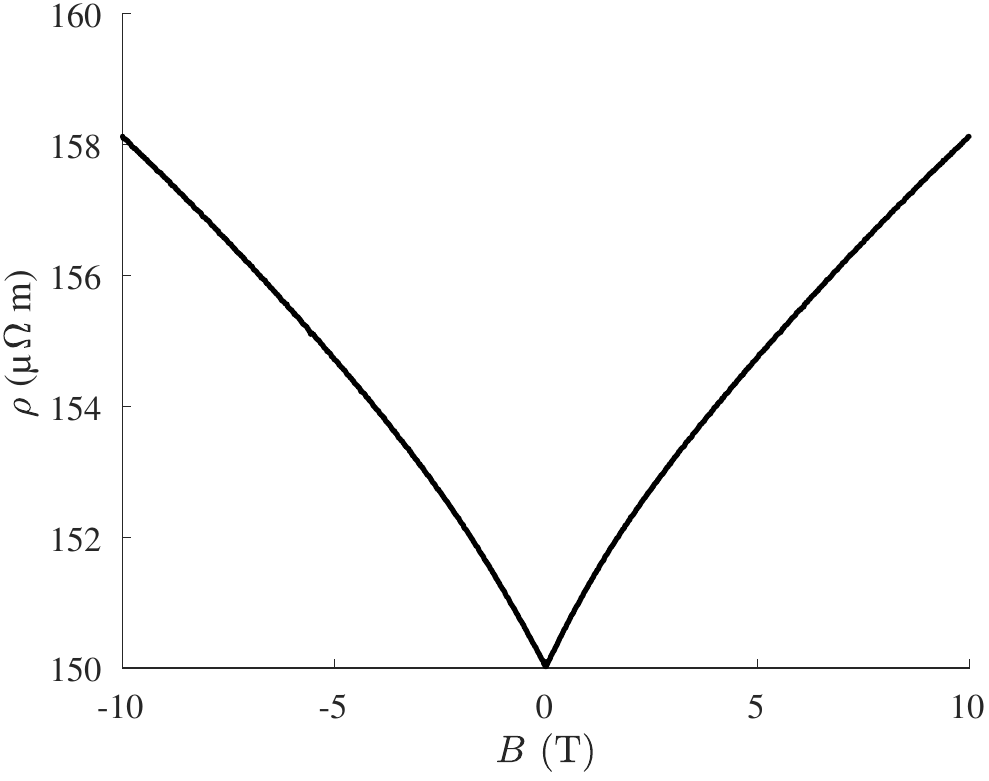}
    \caption{$\rho(B)$ of \AuGex{0.07}{0.93} at \SI{1.8}{\kelvin}. The high field magnetoresistance is approximately linear.}
    \label{fig:Magnetoresponse}
\end{figure}

What is the response of the \AuGe alloy to a magnetic field $B$? As shown in \figref{Magnetoresponse} for $T \gtrsim \SI{1}{\kelvin}$ the films show an approximately linear magnetoresistance at high $B$~\cite{dreizin1973anomalous, stroud1976magnetoresistance, parish2005classical} with weak anti-localization at small $B$. This allows, in principle, for a simple calibration-based read-out of $T$ in the presence of a finite $B$. We also note that at \SI{1.8}{\kelvin} the magnetoresistance is $\approx \SI[per-mode=symbol]{0.5}{\percent\per\tesla}$, implying that the induced error is $\approx \SI[per-mode=symbol]{0.13}{\kelvin\per\tesla}$. Below \SI{1}{\kelvin}, the $B$-response of \AuGe becomes more complex with $\rho(B)$ showing non-monotonic behavior and slow transients similar to that seen in \cite{narayan19, verpoort19}.

\begin{figure}
   \includegraphics[width=\columnwidth]{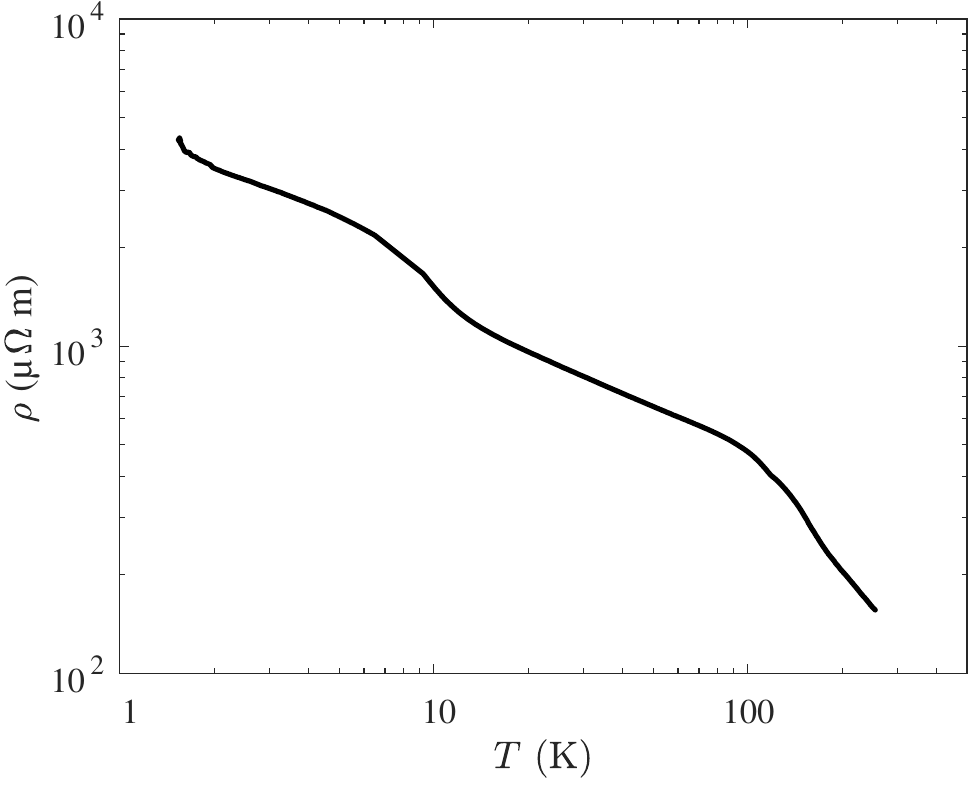}
    \caption{$\rho(T)$ for a \AuGex{0.07}{0.93} film heated at $\approx \SI{250}{\celsius}$ post deposition does not have a logarithmic $T$ dependence but is approximately consistent with a power-law dependence. Furthermore, the absolute values of $\rho$ are significantly larger than in otherwise identical samples heated to $\approx\SI{450}{\celsius}$ (\figref{AG3-1a_RvsT}).}
    \label{fig:AG3-3b_RvsT}
\end{figure}

The post-deposition heat treatment of the \AuGe alloy has a crucial impact on its performance as a thermometer. Figure~\ref{fig:AG3-3b_RvsT} shows $\rho \left(T\right)$ for a sample of \AuGex{0.07}{0.93}, with the material deposited at the same time as for the sample shown in \figref{AG3-1a_RvsT}, but importantly with the only difference being that it was annealed at \SI{250}{\celsius}, which is below the eutectic point $T_{\mathrm{E}}$ of Au-Ge alloys ($\approx$\SI{360}{\celsius})~\cite{okamoto1984}. The sample in \figref{AG3-3b_RvsT} does not show a logarithmic $\rho \left(T\right)$, but instead $\rho(T)$ is closer to a power-law dependence, consistent with a previous work that suggested the use of \AuGe alloys as thermometers, where the films were heated at $T < T_{\mathrm{E}}$~\cite{bethoux1995auge}. Clearly, the overall characteristics are far less satisfactory in terms of how sensitivity and accuracy vary with temperature compared to those shown in \figref{AG3-1a_RvsT}.

The observed logarithmic $\rho(T)$ is clearly correlated to the heat treatment of the films. Cochrane~\textit{et al.}~\cite{cochrane_structural_1975} previously reported logarithmic $\rho(T)$ dependences in \textit{moderately}-disordered materials and explained their findings based on disorder-induced, non-magnetic localized microscopic degrees of freedom, which act as sources for Kondo-type scattering events. These localized degrees of freedom correspond to different configurations of atomic positions that are separated by potential wells and have a finite tunnelling amplitude. Unlike in the case of localized magnetic spins, the resulting eigenstates are not perfectly degenerate, but instead have their energy levels split by $\Delta$. This gives a resistance dependence of
\begin{equation}
 \rho \propto -\log\left[\left(\frac{\kB T}{D}\right)^2 + \left(\frac{\Delta}{D}\right)^2\right]\punc{,}
\label{Disorder_Kondo}
\end{equation}
where $D$ is the electron bandwidth. Consequently, the logarithmic behavior ceases when $\kB T \lesssim \Delta$, and $\rho$ becomes independent of $T$.

The visibly-disordered character of the \AuGe films suggests that the observed logarithmic $\rho(T)$ arises due to similar physics, as indicated by transmission electron micrograph (TEM) images (inset \figref{RvsTAllSamples}) that reveal that the films are phase-separated. Consideration of the phase diagram \cite{okamoto1984} suggests that the two phases are Au-rich (containing $\sim 30\%$ Ge) islands and an almost pure Ge matrix. This is consistent with energy dispersive X-ray spectroscopy measurements. Generally, it is observed that as the fraction of Au is decreased, the logarithmic fit for $\rho\left(T\right)$ becomes better (\figref{RvsTAllSamples}), and this might suggest that the logarithmic conductivity arises dominantly through the Ge matrix. In samples with large contiguous regions of Au-rich material, conduction through the Au-rich regions becomes more important, and causes a correction to the $\rho \left(T\right)$ dependence. We speculate that the Ge matrix is slightly disordered, presumably due to the rapid cooling after the heat treatment, and/or due to the presence of Au impurities that disturb the crystallinity. The flattening of $\rho(T)$ at low $T$ (\figref{AG3-1a_RvsT}) would suggest that $\Delta \approx \SI{0.2}{\kelvin} \cdot \kB$ in the \AuGe films. The observed `shoulder' in $\rho(T)$ at $T \approx \SI{200}{\kelvin}$ is also consistent with the observations in Ref.~\cite{cochrane_structural_1975} and we speculate that these can arise due to higher energy levels of the Kondo-type scattering centre. Notably, over the explored $T$ range we see no indication of a minimum in $\rho(T)$, which is expected at a certain $T$, depending on the concentration of Kondo-type scatterers~\cite{kondo1964minimum}. Separately, we also note that the observed linear magnetoresistance (\figref{Magnetoresponse}) is consistent with the picture of microscopic disorder playing a pivotal role in the transport, as linear magnetoresistance is characteristic of disordered materials~\cite{dreizin1973anomalous, stroud1976magnetoresistance, parish2005classical}.

We note that there are other possible sources of a logarithmic temperature dependence, such as granular systems~\cite{Beloborodov2007} and thin disordered films~\cite{Lee1985}, but these apply to electrical conductivity rather than resistivity as seen here. As shown in Fig.~S4, $\rho \left( T \right)$ clearly fits a $\log \left( T \right) $-dependence more consistently.

\section{Conclusion}

In conclusion we have presented a \AuGe alloy with clear practical advantages for simple $R$-based thermometry from room $T$ down to $\approx \SI{0.2}{\kelvin}$. Principally, the use of a \textit{single sensor} over this $T$ range eliminates the necessity of matching calibrations of different thermometers at the limits of their applicability, while also reducing costs towards multiple sensors and the associated wiring and measurement apparatus. Furthermore, the reasonably low values of $R$ over the entire $T$ range alleviates the need for apparatus that can measure $R$ with high precision and without appreciable self-heating. The possibility of lithographically patterning the alloy suggests that \si{\micro\metre}-scale thermometers can be deposited directly onto substrates of choice, allowing for thermal and thermoelectric measurements on small samples and at low $T$. The accuracy of the logarithmic $T$-dependence over a wide $T$ range allows for a simple 2 point calibration. We find that the characteristic logarithmic $\rho(T)$ behavior of the alloy is linked to its composition and the precise conditions of preparation, in particular annealing above the eutectic temperature. We suggest that the mechanism driving the logarithmic resistance dependence on temperature is the same as that reported previously in moderately-disordered metallic alloys, namely where the disorder induces Kondo-like scattering centres. Strikingly, however, the $T$ range over which we observe the logarithmic behavior is significantly larger than in any previous report we are aware of, with the anticipated `upturn' in resistance not occurring up to room temperature.

\section{Acknowledgements}

J.R.A.D., P.C.V., G.J.C., and V.N. acknowledge funding from the Engineering and Physical Sciences Research Council (EPSRC), UK. G.J.C. and S.E.R. acknowledge funding from the Royal Society, UK. J.F.O. would like to thank the Brazilian Agencie CNPq. A.D. and S.K-N. acknowledge financial support through an ERC Starting Grant (Grant No. ERC-2014-STG-639526, NANOGEN). We thank J. J. Rickard for support with TEM.

\bibliography{references}

\end{document}


\title{Au-Ge alloys for wide-range low-temperature on-chip thermometry: Supplemental material}
\author{J. R. A. Dann}
\email{jrad3@cam.ac.uk}
\affiliation{Cavendish Laboratory, University of Cambridge, J.J. Thomson Avenue, Cambridge CB3 0HE, UK}
\author{P. C. Verpoort}
\affiliation{Cavendish Laboratory, University of Cambridge, J.J. Thomson Avenue, Cambridge CB3 0HE, UK}
\author{J. Ferreira de Oliveira}
\affiliation{Cavendish Laboratory, University of Cambridge, J.J. Thomson Avenue, Cambridge CB3 0HE, UK}
\affiliation{Centro Brasileiro de Pesquisas Físicas, Rua Dr Xavier Sigaud 150, Rio de Janeiro, 22290-180, Brazil}
\author{S. E. Rowley}
\affiliation{Cavendish Laboratory, University of Cambridge, J.J. Thomson Avenue, Cambridge CB3 0HE, UK}
\affiliation{Centro Brasileiro de Pesquisas Físicas, Rua Dr Xavier Sigaud 150, Rio de Janeiro, 22290-180, Brazil}
\author{A. Datta}
\affiliation{Department of Materials Science, University of Cambridge, 27 Charles Babbage Road, Cambridge CB3 0FS, UK}
\author{S. Kar-Narayan}
\affiliation{Department of Materials Science, University of Cambridge, 27 Charles Babbage Road, Cambridge CB3 0FS, UK}
\author{C. J. B. Ford}
\affiliation{Cavendish Laboratory, University of Cambridge, J.J. Thomson Avenue, Cambridge CB3 0HE, UK}
\author{G. J. Conduit}
\affiliation{Cavendish Laboratory, University of Cambridge, J.J. Thomson Avenue, Cambridge CB3 0HE, UK}
\author{V. Narayan}
\email{vn237@cam.ac.uk}
\affiliation{Cavendish Laboratory, University of Cambridge, J.J. Thomson Avenue, Cambridge CB3 0HE, UK}
\date{\today}

\maketitle

\begin{figure}
	\centering
	\includegraphics[width=0.5\textwidth]{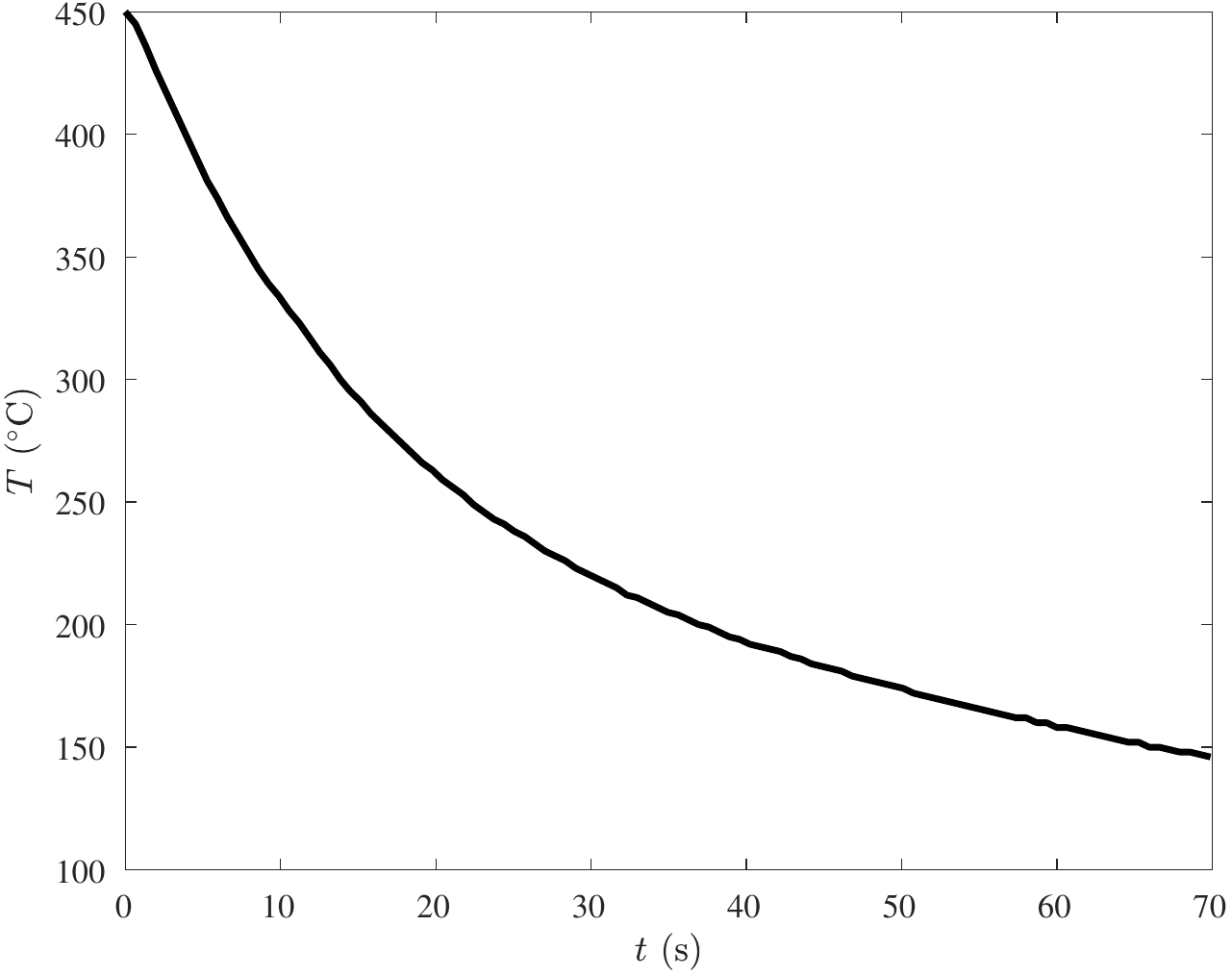}
	\caption{A typical cooling curve for the heat treatment of the \AuGe films.}
\end{figure}

\begin{figure}[h]
	\includegraphics[width=\textwidth]{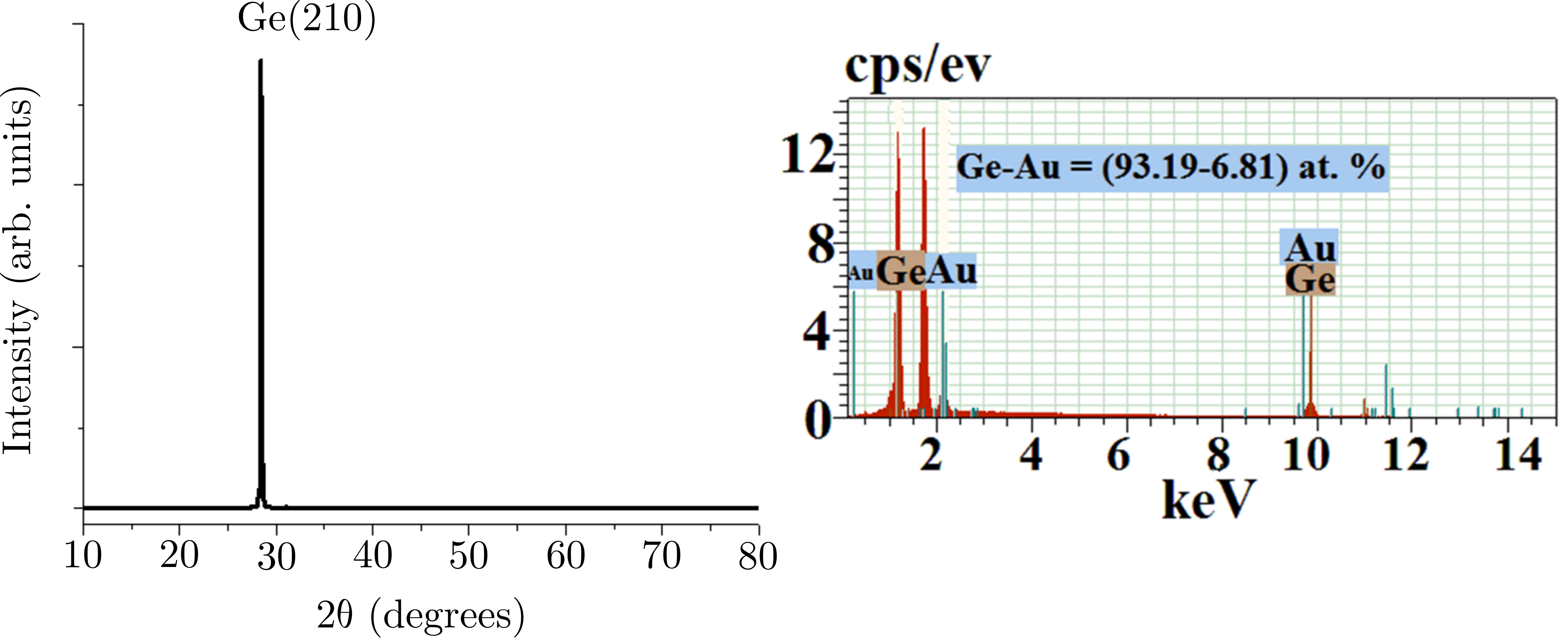}
	\caption{Despite the grainy structure evidenced in Fig.~3, XRD images (left panel) show a clear Ge(210) peak suggesting the Ge-matrix to have crystalline order. The Au concentration of 7\% is below the XRD resolution, but is confirmed using energy dispersive X-ray spectroscopy (right panel).}
\end{figure}

\begin{figure}
	\includegraphics[width=0.5\textwidth]{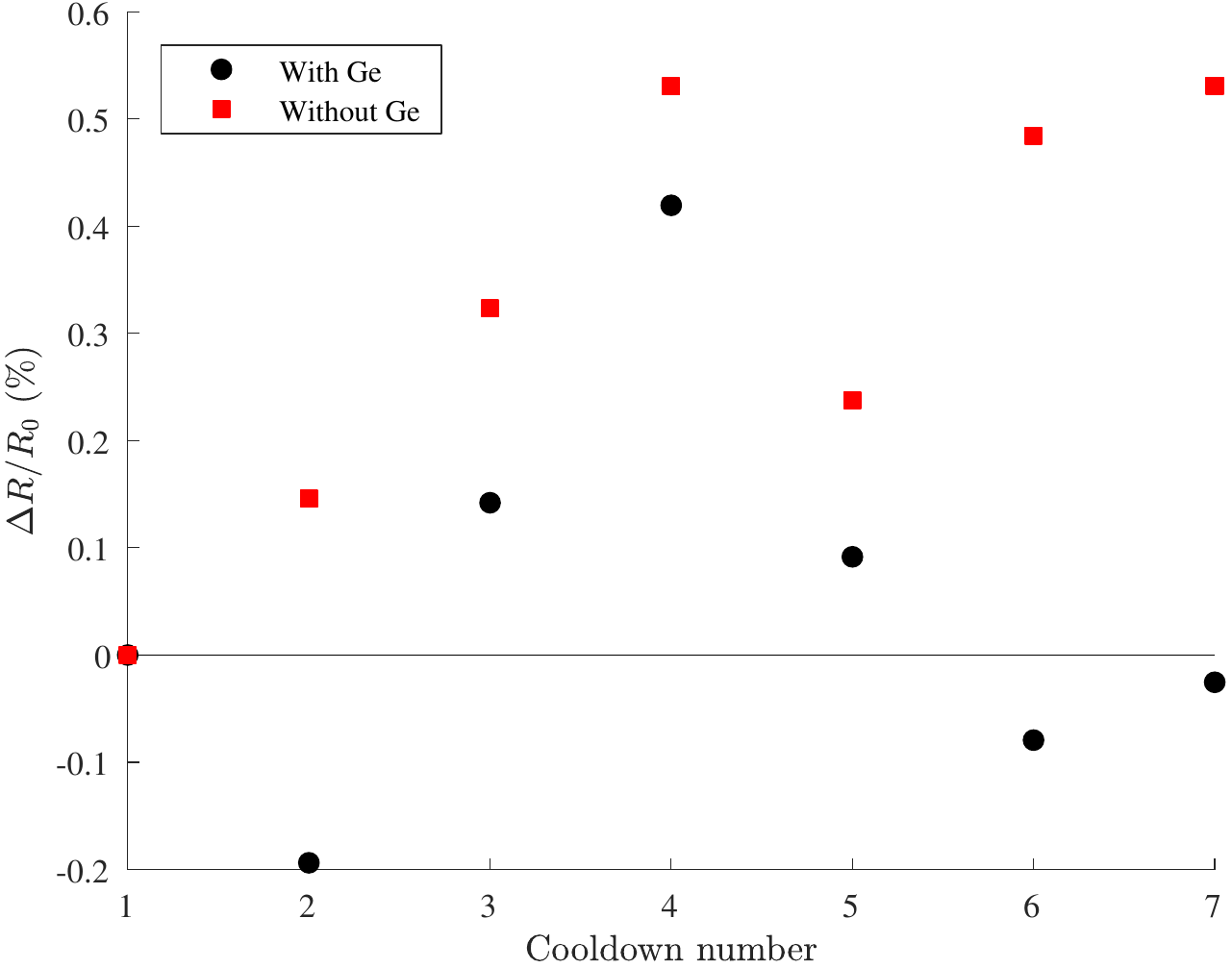}
	\caption{Variation in resistance, $R$, between cooldowns at \SI{4}{\kelvin} for \AuGex{0.07}{0.93} films with (black circles) and without (red squares) a Ge capping layer. $\Delta R$ is the change in resistance from the first cooldown and $R_0$ is the resistance during the first cooldown.}
\end{figure}

\begin{figure}
	\includegraphics[width=\textwidth]{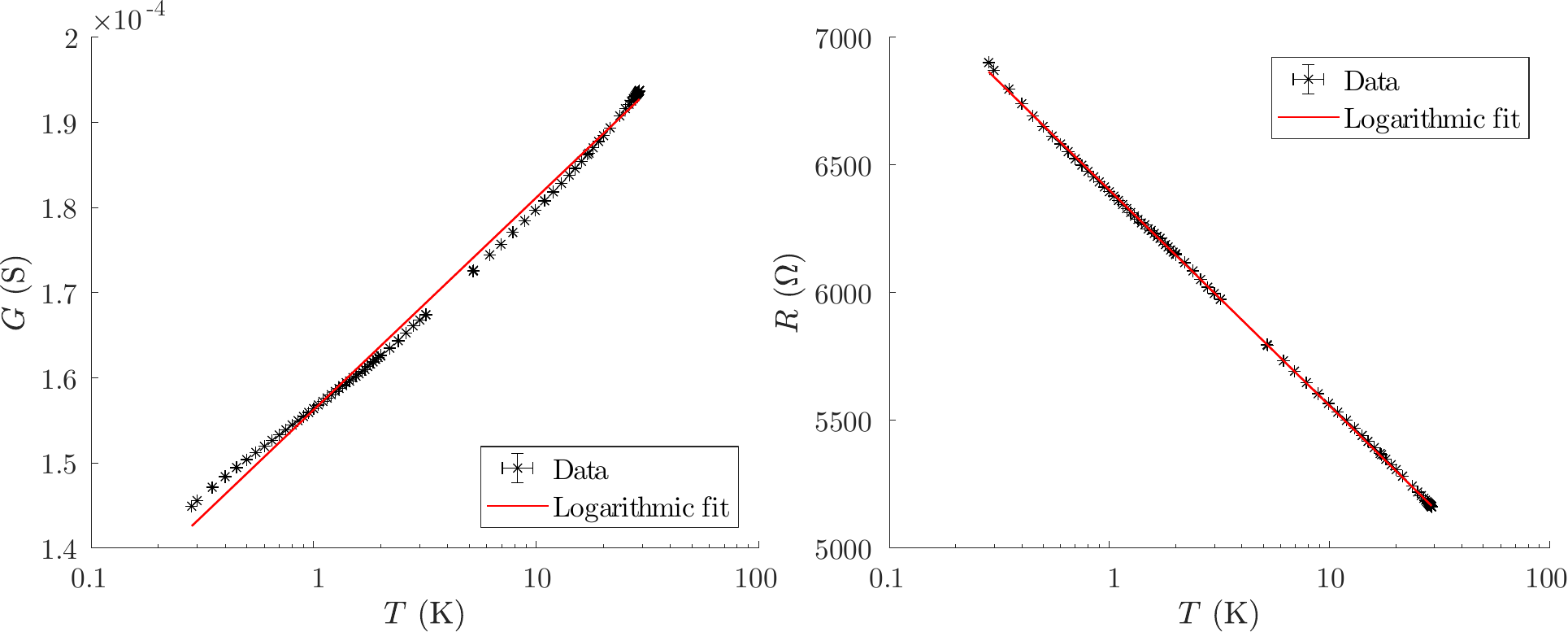}
	\caption{Graphs showing conductance (left panel) and resistance (right panel) against temperature for a \AuGex{0.07}{0.93} sample, with logarithmic fits. The resistance clearly fits a logarithmic temperature dependence much better.}
\end{figure}